\documentclass[aps,prd,twocolumn,amsmath,amssymb,amsfonts,nofootinbib,superscriptaddress,altaffilletter]{revtex4-2}

\usepackage{graphicx}
\usepackage{hyperref}
\hypersetup{
  bookmarksopen=true
}
\usepackage{ulem}
\usepackage{amssymb}
\usepackage{amsmath}
\usepackage{color}
\usepackage{supertabular}
\usepackage{dcolumn}
\usepackage[mathscr]{eucal}
\usepackage{mathrsfs}
\usepackage{siunitx}

\def\GPS{1253800000}
\def\GPSUTC{2019 Sep 29 13:46:22 UTC}

\def\sci#1#2{#1\times10^{#2}}

\begin{document}

\title{Early release of low-frequency atlas of continuous gravitational waves}

\author{Vladimir Dergachev}
\email{vladimir.dergachev@aei.mpg.de}
\affiliation{Max Planck Institute for Gravitational Physics (Albert Einstein Institute), Callinstrasse 38, 30167 Hannover, Germany}

\author{Maria Alessandra Papa}
\email{maria.alessandra.papa@aei.mpg.de}
\affiliation{Max Planck Institute for Gravitational Physics (Albert Einstein Institute), Callinstrasse 38, 30167 Hannover, Germany}
\affiliation{Leibniz Universit\"at Hannover, D-30167 Hannover, Germany}

\begin{abstract}
We present the public release of the low-frequency atlas of continuous gravitational waves, covering signals with frequencies from 20\,Hz to 200\,Hz and frequency derivatives from $\sci{-5}{-11}$ to $\sci{5}{-11}$\,Hz/s. Compared to the previous atlas releases, this version demonstrates significant improvements in sensitivity and sky resolution. In the most sensitive region even the worst-case upper limits on gravitational wave strain are below $10^{-25}$. The atlas data is being released ahead of the completion of the full follow-up analysis. 
\end{abstract}

\maketitle

\section{Introduction}

The first continuous gravitational wave atlas was released several years ago \cite{o3a_atlas1} and it covered signals with frequencies between 500-1000\,Hz and frequency derivatives from $\sci{-5}{-11}$ to $\sci{5}{-11}$\,Hz/s. Since then, we produced new atlases that progressively expanded parameter space coverage \cite{o3a_atlas1, o3a_atlas2, o3a_atlas3}.

\begin{figure}[htbp]
\includegraphics[width=3.3in]{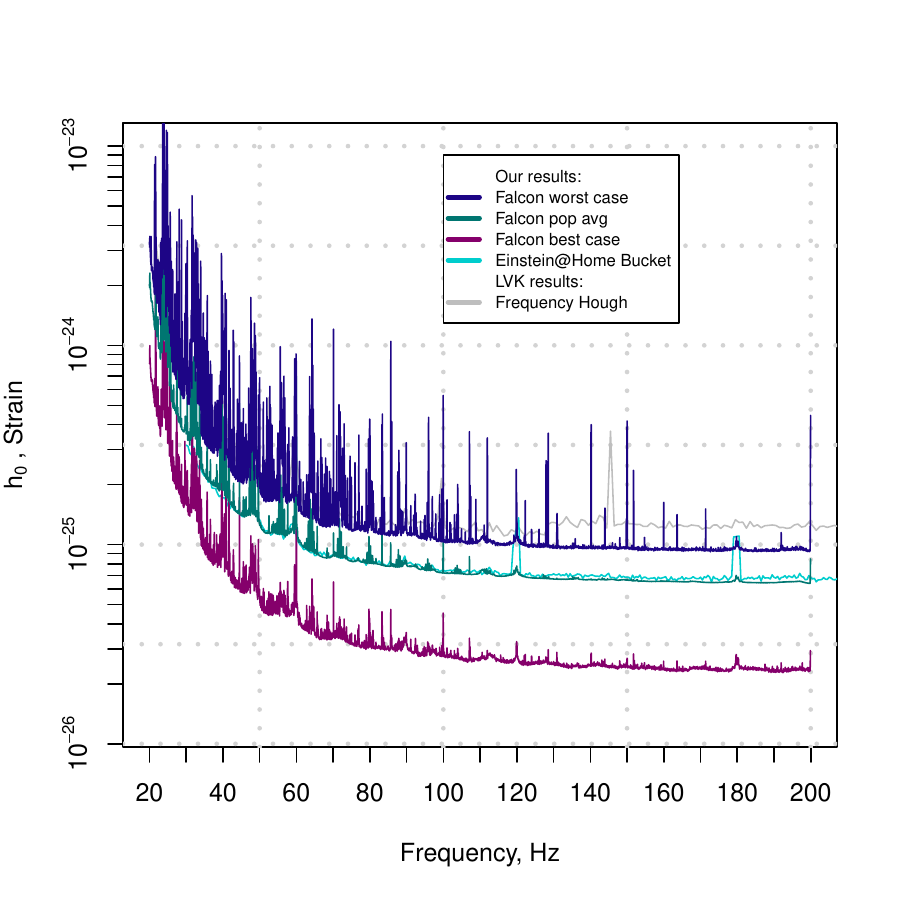}
\caption[Upper limits]{
\label{fig_amplitudeULs}
Gravitational wave intrinsic amplitude $h_0$ upper limits at 95\% confidence as a function of signal frequency. The upper limits reflect the sensitivity of the search. We also plot latest LIGO/Virgo and Einstein@Home all-sky population average results \cite{lvc_O3_allsky2, EatHO3Bucket, o3a_atlas2} and Falcon population average proxy.
}
\end{figure}

With this paper, we make the low-frequency atlas publicly available to improve coverage in the low frequency regime. The sky resolution has been enhanced by a factor of $6$ compared to \cite{o3a_atlas3}, and the use of the full O3 data \cite{O3aDataSet, O3bDataSet} combined with a longer coherence length has improved the sensitivity of the search. In fact, the worst-case all-sky upper limits have decreased by 30\%, and for the first time are below $10^{-25}$ at the most sensitive frequencies (Figure \ref{fig_amplitudeULs}).

Our atlas provides upper limits on continuous gravitational waves based on measurements of power in LIGO \cite{aligo} data. Additionally, we include signal-to-noise ratio (SNR) peaks and their associated parameters.

\section{Signal model}

Continuous gravitational waves are expected from rapidly rotating neutron stars with sustained non-axisymmetric deformation $\varepsilon$. The intrinsic amplitude of the gravitational wave signal at a distance $d$ from the source is proportional to the deformation, increases with the square of the signal frequency $f$ and decreases with the inverse of the distance to the source $d$:
\begin{equation}
h_0=\frac{4\pi^2G I_{zz} f^2 \varepsilon}{c^4 d},
\label{eq:epsilon}
\end{equation}
where $I_{zz}=10^{38}$\,kg\,m$^2$ is the moment of inertia of the star with respect to the principal axis aligned with the rotation axis.  Gravitational waves carry away energy, gradually slowing down the pulsar's rotation. Most known radio pulsars have frequency derivatives below $\pm \sci{5}{-11}$\,Hz/s.

To compute the atlas data we assume a source which emits gravitational waves with the following frequency evolution in the Solar System barycenter
\begin{equation}
f(t)=f_0+(t-t_0)f_1+(t-t_0)^2f_2/2.
\label{eq:freqEvolution}
\end{equation}
Here source frequency $f_0$ covers frequencies from $20$ to $200$\,Hz, source frequency derivative is constrained by $|f_1|\le \sci{5}{-11}$\,Hz/s and source second frequency derivative is within $|f_2|\le 10^{-20}$\,Hz/s$^2$.
We chose epoch $t_0=\GPS$\,s\,\,GPS (\GPSUTC), which is in the middle of our data set.

Higher order frequency derivatives are not explicitly searched over, instead relying on intrinsic robustness of a search with coherence lengths of 2\,days and using one year of data.

The signal model of equation \ref{eq:freqEvolution} and the parameter space  choice were optimized for isolated rotating neutron stars, but apply more generally, in particular neutron stars in wide-period binary orbits \cite{Singh:2019han} and boson condensate sources \cite{boson1, boson2, discovering_axions, Zhu:2020tht}.

\section{Atlas data}

The atlas \cite{data} partitions the parameter space by frequency and sky position. The frequency space is divided into 50\,mHz frequency bands. For each frequency band, we construct a grid of sky locations. The sky points are sufficiently dense that the upper limit values associated to the point hold for all sky positions in the Voronoi domain associated with the point. The Voronoi domain is constructed using the angular distance on the celestial sphere.

We set different types of upper limits, established based on different assumptions on the polarization of incoming signals. The worst-case upper limit is computed as maximum over all polarizations and is dominated by linearly polarized signals. Circularly polarized signals yield the most stringent upper limits (lower curve in Figure \ref{fig_amplitudeULs}), typically more than a factor of $2$ lower than worst-case upper limits, because circularly polarized signals couple best with the detector. Upper limits for signals of arbitrary polarizations can be estimated using the sky-position-dependent coefficients provided in the atlas\cite{functional_upper_limits}.

All upper limits are derived using the universal statistics algorithm \cite{universal_statistics} and are strict frequentist 95\% confidence level upper limits. The 95\% is actually a design target, and actual values are expected to have higher confidence depending on the underlying noise distribution.

The SNR values included in the atlas data represent  peak values over the Voronoi domain associated with the given sky point and over the given frequency band. We also provide the value of the frequency and the polarization of the highest SNR result. 

With this atlas we also introduce an estimate of the signal strength necessary for our search to ``lock-on" to a signal from a given sky-position, i.e. to detect the signal and correctly estimate its frequency and polarization. This is the  ``location-specific polarization-average lock-level proxy".

\begin{table}[htbp]
\begin{center}
\begin{tabular}{l D{.}{.}{3.5} r r r r r r }
\hline
 \# & \multicolumn{1}{c}{$f$} & \multicolumn{1}{c}{$\dot{f}$} &  SNR & UL/$h_0$ & $h_0$/LL &$\Delta f$ & In \\
& \multicolumn{1}{c}{Hz} & Hz/s & (atlas) & \% & \% & mHz & \\
\hline
\hline
%
3 &     108.85716 &      -1.46e-17 &     34.1 &       126.7 &      189.4 &      0.031 &    Yes \\
5 &     52.80832 &      -4.03e-18 &     259.9 &      135.2 &       364.6 &      -0.005 &  Yes \\
6 &     145.33976 &       -6.73e-09 &     8.2 &      19.8 &      638.0 &      -30.505 &      No \\
8 &     189.96498 &       -8.65e-09 &     8.3 &      58.8 &      215.0 &      0.832 &     No \\
10 &    26.33144 &       -8.5e-11 &      51.2 &      86.4 &      120.5 &      -23.734 &     No \\
11 &    31.42470 &      -5.07e-13 &     41.7 &      245.4 &      87.4 &      -10.257 &      Yes \\
12 &    37.70750 &        -6.25e-09 &     9.7 &      106.4 &      133.3 &      47.890 &      No \\

\hline
\end{tabular}

\caption[Hardware injections]{This table shows the frequency parameters of the hardware-injected continuous wave signals and how they appear in the atlas data.  We show all the hardware injections within 20-200\,Hz range, including those outside of our search space, as indicated by the ``In'' column. We show our upper limit UL at the injection sky position and polarization parameter values, and how it compares with the injection amplitude $h_0$: if the upper limits are correct, UL/$h_0$ has to be $> 100$\%.
$h_0$/LL shows the ratio of signal strain to the lock level, and we expect to detect signals when this ratio is $> 100\%$. This is the case for the first two hardware-injected signals in range, but not for the third, as it can be seen from the large mismatch difference $\Delta f$  between the recovered frequency and the signal frequency. 
For detectable signals in our parameter space we expect the $\Delta f$ to be within $0.5$\,mHz.  We use the reference time (GPS epoch) $t_0=\GPS$\,s (\GPSUTC). The script {\tt spatial\_index\_example2.R} provided with the atlas shows how to obtain this data from the atlas.}
\label{tab:injections}
\end{center}
\end{table}

\section{Population average proxies}
\label{sec:proxies}

The location-specific polarization-average lock-level proxy is defined as the signal strain level at which the frequency and polarization parameters of the SNR peak in the corresponding Voronoi domain (sky region around a sky-grid point and 50 mHz frequency band) are correctly reported in the atlas for 95\% of signals. The primary intended use for the proxy is to assess when a limited parameter-space follow-up of SNR-peak parameters has merit, or when the entire Voronoi domain needs to be reanalyzed with a more sensitive search in the given frequency band.

\begin{figure}[htbp]
\includegraphics[width=3.3in]{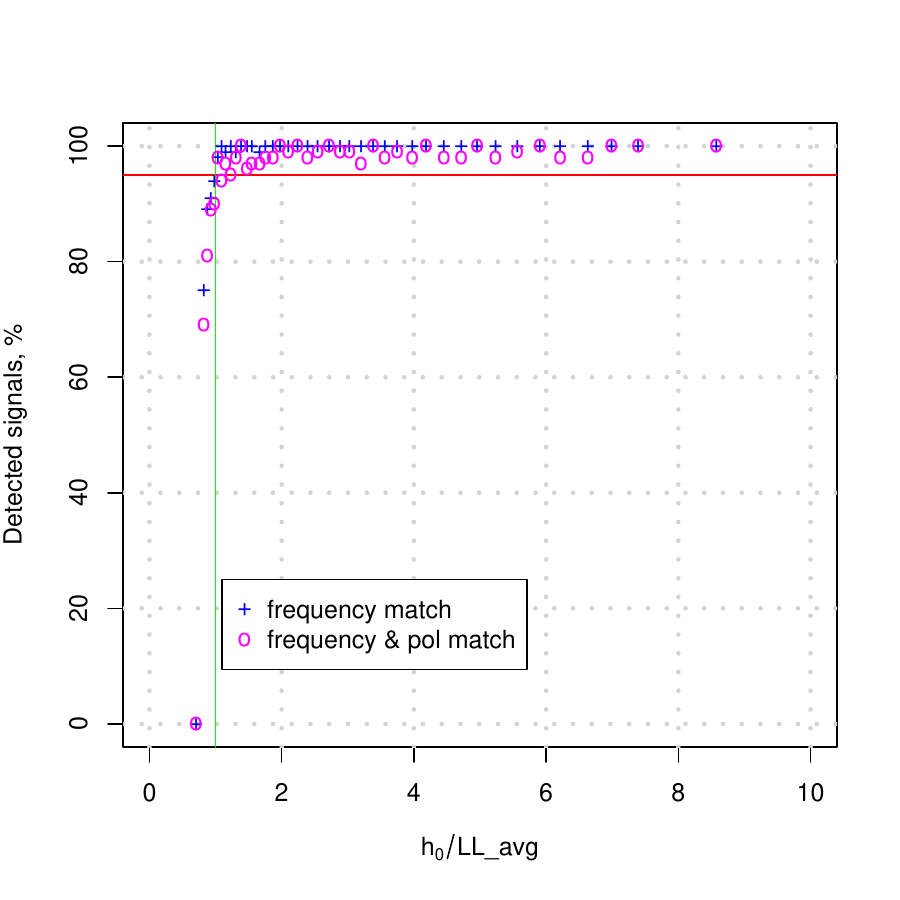}
\caption[Signal recovery]{
\label{fig_signal_recovery}
Signal recovery versus the ratio of signal strain to lock-level estimated on data with the test signals. The red line marks the $95$\% threshold. The blue crosses show the percentage of points where the SNR peak frequency was withing $0.2$\,mHz of the injected value. Given the $50$\,mHz frequency band there is a 0.4\% chance of false coincidence. The magenta circles show the percentage of points where $\iota$ was within $0.7/\sin(\iota)$ of injected value and $\psi$ was also within $0.7$ of injected value modulo $\pi/2$. Such wide tolerances are due to shoulders in $\iota$ and $\psi$ recovery (see Figures~\ref{fig_iota_recovery} and \ref{fig_psi_recovery}).
}
\end{figure}

Our location-specific polarization-average lock-level proxies are computed for each atlas entry from an estimate of the background noise and are insensitive to the presence of plausible strength signals. In our injection study 99\% of the test signals (of Figure~\ref{fig_signal_recovery}) change the lock level by less than 1\%. 

\begin{figure*}[htbp]
\includegraphics[width=6.6in]{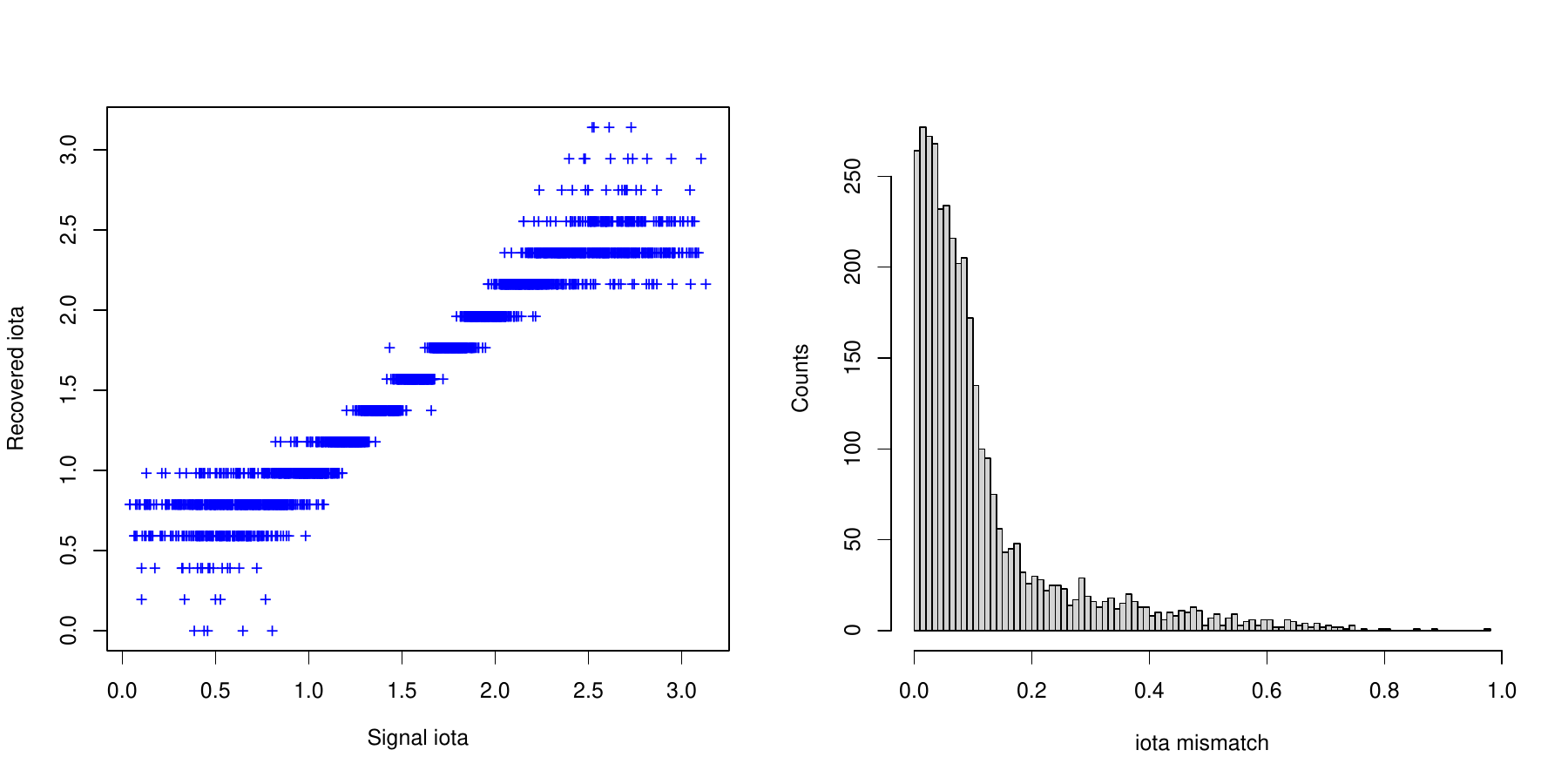}
\caption[Iota recovery]{
\label{fig_iota_recovery}
Recovery of the inclination angle $\iota$. We plot only points corresponding to signals with strains at or above the population average lock-level. The higher uncertainty for nearly circular polarized signals results in wide shoulders in the histogram.
}
\end{figure*}

\begin{figure*}[htbp]
\includegraphics[width=6.6in]{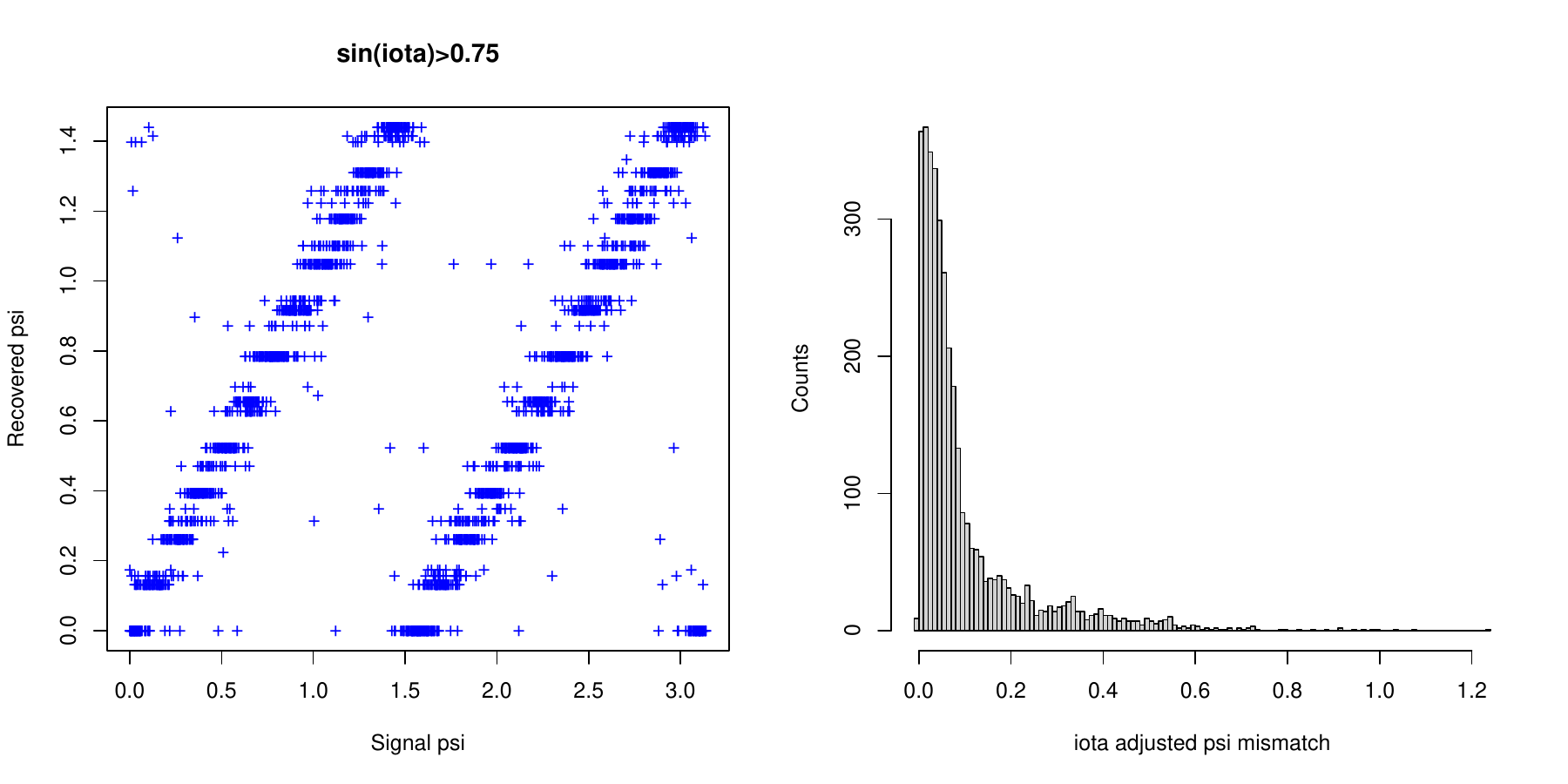}
\caption[Psi recovery]{
\label{fig_psi_recovery}
Recovery of  the polarization angle $\psi$ from the SNR peaks. We plot only points corresponding to injected strain at or above the polarization-average lock-level. The left plot restricts data to nearly linearly polarized signals. The right plot includes all points, but multiplies the difference between injected and recovered $\psi$ values (modulo $\pi/2$) by $\sin(\iota)$. This way the mismatch in $\psi$ is deweighted as $\iota$ approaches circular polarizations $0$ or $\pi$ (the ``poles'' of the polarization sphere).}
\end{figure*}

We validate this proxy using test signals distributed uniformly on the sky and with frequencies in the 100-200\,Hz band. For each test signal, we consider 
the parameters of the SNR peak at the sky-point closest to the position of the test signal and in the relevant 50 mHz frequency band. Figure~\ref{fig_signal_recovery} shows the percentage of signals with SNR peak frequency  within $0.5$\,mHz of the actual value, as well as the  percentage of signals with both frequency and polarization correctly recovered, as a function of signal strength. We verify that when the signal strength is at the lock-level, the recovered fraction of signals across a population of signals with arbitrary polarization is at the 95\% level. Figure~\ref{fig_signal_recovery}  also shows that frequencies are recovered most accurately, while polarization recovery is less precise. This is because polarization reconstruction relies heavily on amplitude modulation, introducing significant  degeneracy, especially for nearly circularly polarized signals (Figures~\ref{fig_iota_recovery} and \ref{fig_psi_recovery}).

Location-specific lock-level proxies can be used to estimate the sensitivity of our search for a population of signals coming from any direction in the sky, and this can in turn be used to generate an estimate of the so-called population-average upper limits over the whole sky, in the absence of signals. This is useful because most all-sky searches \cite{lvc_O3_allsky2, EatHO3a, EatHO3Bucket,EatHO3HF} report results population-average upper limits. These are conceptually simple to implement by simulating search results from populations of signals and comparing with the observation results, until the signal amplitude that yields the 95\% quantile is identified. However, practical challenges may arise, and the computational cost of these estimates is generally not trivial.
The Falcon searches, like the one presented here, and its predecessor the Powerflux search, have used a different approach, establishing upper limits for fixed polarizations using the universal statistics \cite{universal_statistics}, which do not suffer from the same limitations. However, this can make comparison with the results of other pipelines more difficult.

With the lock-level proxies in hand, we provide an estimate of the population-average all-sky upper limits in the absence of signals, by taking the 95\% quantiles of the location-specific lock-level proxies, resulting in a 90\% confidence level.  The resulting curve is shown in Figure \ref{fig_amplitudeULs}, and it can be used for pipeline sensitivity comparisons. Its physical meaning should be interpreted as a level of strain at which 90\% of signals are expected to have correct SNR peak parameters in the atlas, and not as actual upper limit results. Computationally expensive test-signal recovery simulations would be required to determine the actual confidence associated with the values in each band.

\begin{table}[htbp]
\begin{center}
\begin{tabular}{r r r r r}
\hline
 Band start & Band stop & SNR &  Frequency & Frequency  \\
\multicolumn{1}{c}{Hz} & \multicolumn{1}{c}{Hz} & all-sky & lock, \% & \& pol lock, \% \\
\hline
\hline
100.003 & 100.054 & 98.2 & 64.0 & 61.0  \\
100.053 & 100.104 & 17.8 & 99.0 & 96.5  \\
100.103 & 100.154 & 10.6 & 99.5 & 98.0  \\
100.153 & 100.204 & 10.3 & 99.5 & 94.0  \\
100.203 & 100.254 & 10.2 & 99.0 & 98.0  \\
100.253 & 100.304 & 12.3 & 99.5 & 97.5  \\
100.303 & 100.354 & 11.3 & 99.5 & 98.0  \\
100.353 & 100.404 & 10.9 & 99.5 & 97.0  \\
100.403 & 100.454 & 10.3 & 100.0 & 97.5  \\
100.453 & 100.504 & 10.5 & 97.0 & 92.5  \\
100.503 & 100.554 & 10.4 & 98.0 & 91.5  \\
100.553 & 100.604 & 10.3 & 99.0 & 97.0  \\
100.603 & 100.654 & 10.4 & 99.5 & 95.5  \\
100.653 & 100.704 & 10.6 & 99.0 & 97.0  \\
100.703 & 100.754 & 10.5 & 100.0 & 95.0  \\
100.753 & 100.804 & 10.6 & 99.5 & 95.5  \\
100.803 & 100.854 & 13.5 & 97.5 & 93.0  \\
100.853 & 100.904 & 10.5 & 99.5 & 95.5  \\
100.903 & 100.954 & 10.4 & 98.0 & 96.0  \\
100.953 & 101.004 & 10.7 & 100.0 & 96.0  \\
\hline
\end{tabular}

\caption[Spot check]{This table shows how signals are recovered in the 100-101\,Hz frequency range and it is an illustrative spot-check of the all-sky population average upper limit estimate. 
In each band we consider 200 test-signals with strain equal to the all-sky population-average upper limit estimate and the other parameters distributed randomly in the search parameter space. We add each signal to the data and carry out the search. Columns 4 and 5 show the percentage of test-signals for which the frequency (column 4) and frequency+polarization (column 5) are correctly recovered in the test signal parameters Voronoi domain. For signals with arbitrary sky position we expect both these numbers, and in particular that of column 5,  to be at least 90\% , if the all-sky population-average upper limit estimate is correct. 
Indeed we see that in most frequency bands this is the case. In the band containing a strong disturbance, such as the first band of the table, the SNR peak is due to the disturbance and not representative of a signal, so ``signal-lock" at the proxy level is not achieved as effectively. Column 3 shows the highest SNR value from each band in the atlas results. Large value indicate a loud signal, astrophysical or not.}
\label{tab_spot_check}
\end{center}
\end{table}

\begin{figure*}[htbp]
\includegraphics[width=6.6in]{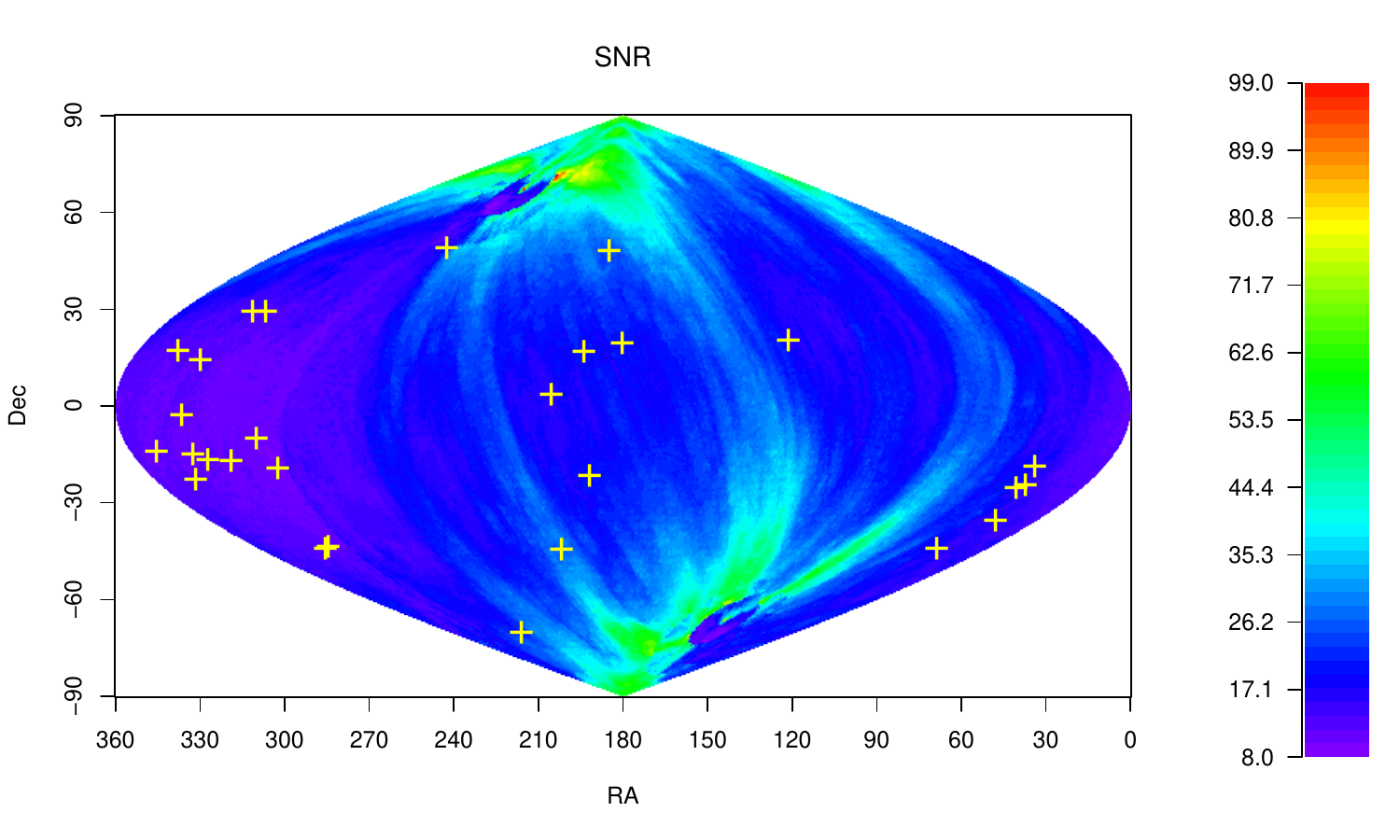}
\caption[Iota recovery]{
\label{fig_skyband100}
This figure shows how SNR varies over sky position in the band 100.003-100.054\,Hz. We observe strong contamination from instrumental artifacts near 100\,Hz. The yellow plus signs denote locations of injections with $|\cos(\iota)|<0.5$ with frequency lock. The injections group in areas with smaller SNR away from contaminated areas.
}
\end{figure*}

We have performed spot-checks of the all-sky population average upper limit estimates, the results of which are shown in Table~\ref{tab_spot_check}. Each frequency band was tested with 200 simulated signals having strain equal to all-sky population average proxy and the rest of parameters distributed randomly. We observe that lock rates are above the design value of 90\% for all bands except 100.003-100.054\,Hz, which is highly contaminated by detector artifacts (Figure \ref{fig_skyband100}).

Figure~\ref{fig_skyband100} shows that nearly linearly-polarized injections fail to achieve frequency lock in contaminated regions with high-SNR from detector artifacts. This is expected behaviour because the lock-level is a measure of background noise and is constructed to be insensitive to any signals in the band - whether astrophysical or technical.

More generally, there is a natural variation in detector sensitivity with sky position. Thus the rate of test-signal recovery based on any population average proxy, regardless of how it is constructed, will vary with the position of the source in the sky. This again highlights the dependence of population average proxy on the population it was designed for, in contrast to frequentist upper limits designed with universal statistics \cite{universal_statistics}, which apply to any subset of parameter space.

\section{Conclusions}
We have conducted a wide-band search of the full O3 data covering frequencies from 20 to 200\,Hz, which is the most sensitive all-sky search to date. The results have been compiled into an atlas, which we release while the final analysis is ongoing. We introduce population average lock-level proxies for every sky location in the atlas data. The atlas includes several examples, written in {\tt R}, that can be a starting point for future searches.

\begin{acknowledgments}
The authors thank the scientists, engineers and technicians of LIGO, whose hard work and dedication produced the data that made this search possible.

The search was performed on the ATLAS cluster at AEI Hannover. We thank Bruce Allen, Carsten Aulbert and Henning Fehrmann for their support.

This research has made use of data or software obtained from the Gravitational Wave Open Science Center (gw-openscience.org), a service of LIGO Laboratory, the LIGO Scientific Collaboration, the Virgo Collaboration, and KAGRA. 
\end{acknowledgments}

\end{document}